\documentstyle[sprocl,epsf]{article}

\def\be{\begin{equation}}
\def\ee{\end{equation}}
\def\bea{\begin{eqnarray}}
\def\eea{\end{eqnarray}}

\begin{document}

\noindent
\thispagestyle{empty}
\renewcommand{\thefootnote}{\fnsymbol{footnote}}
\begin{flushright}
{\bf CERN-TH/2000-143}\\
{\bf hep-ph/0008101}\\
{\bf August 2000}\\
\end{flushright}
\vspace{1.cm}
\begin{center}
  \begin{large}\bf
Non-Relativistic Effective Field Theory for \\[2mm]
       Perturbative Heavy Quark--Antiquark Systems
  \end{large}
  \vspace{2.2cm}

\begin{large}
 A.~H.~Hoang 
\end{large}

\vspace{.2cm}
\begin{center}
\begin{it}
Theory Division, CERN,\\
   CH-1211 Geneva 23, Switzerland
\end{it} 
\end{center}

  \vspace{2cm}
  {\bf Abstract}\\
\vspace{0.3cm}
%
\noindent
\begin{minipage}{11.0cm}
\begin{small}
Recent developments for an effective theory for non-relativistic
perturbative quark--antiquark systems are reviewed and some
applications are discussed.
%
%
\end{small}
\end{minipage}

\vspace{2cm}
{\it Invited talk given at the 5th Workshop on QCD (QCD 2000),
Villefranche-sur-Mer, France, 3-7 January 2000 }
\end{center}
\vspace{1.9cm}
{\bf CERN-TH/2000-143}\\
{\bf August 2000}
%
%
%
\newpage

\title{Non-Relativistic Effective Field Theory for \\
       Perturbative Heavy Quark--Antiquark Systems\footnote{
This write-up was updated for some relevant
publications that have appeared after the time of the conference.
}}

\author{A. H. HOANG}

\address{Theory Division, CERN\\
CH-1211 Geneva 23, Switzerland\\
E-mail: andre.hoang@cern.ch}

\maketitle\abstracts{ 
Recent developments for an effective theory for non-relativistic
perturbative quark-antiquark systems are reviewed and some
applications are discussed.
}

\section{Introduction}

The study of non-relativistic two-body systems consisting of a heavy
quark--antiquark ($Q\bar Q$) pair has a long tradition since the
discovery of the $J/\psi$ and the work of Appelquist and
Politzer~\cite{Appelquist1}, which had shown that, owing to asymptotic
freedom, non-relativistic quantum mechanics should apply to heavy
$Q\bar Q$ systems. Thus, heavy $Q\bar Q$ systems share many features
of the positronium and are an ideal tool for studying the interplay of
perturbative and non-perturbative physics. It is clear that
non-perturbative effects become suppressed more efficiently for
increasing heavy quark mass. To get a more quantitative feeling
about this, it is helpful to consider the scales that are relevant to
the dynamics of a positronium-like $Q\bar Q$ pair with relative
velocity $v\ll 1$, neglecting for now
the influence of non-perturbative effects. There are three: the mass,
$m_Q$ (relevant to processes with high virtuality such as
annihilation), $m_Q v$ (which sets the size of the system and
the relative momentum), and $m_Q v^2$ (which is of order of the binding 
energy). To assess the importance of non-perturbative effects, one can
compare these scales with the typical hadronic scale $\Lambda_{\rm QCD}$.
The tools needed to describe the $Q\bar Q$ pair depend on the relation
of $\Lambda_{\rm QCD}$ to the three scales just mentioned.

In this talk I consider heavy $Q\bar Q$ systems where 
$m_Q\gg m_Q v\gg m_Q v^2 \gg \Lambda_{\rm QCD}$, and I call such
systems ``perturbative''\cite{Brambilla1}. 
In fact, only a top--antitop ($t\bar t$) pair can
clearly satisfy this condition, and it satisfies it even for
arbitrarily small  
velocities because of the large top decay rate. (I will come back to
this point in Sec.~\ref{sectionttbar}.) For a bottom--antibottom
($b\bar b$) pair, 
the condition is, if at all, only valid for the bottomonium ground
state, whereas for a charm--anticharm pair it can certainly never be
satisfied. For $b\bar b$ pairs, on the other hand, one can construct
quantities, such as moments of the total cross section of hadrons
containing a bottom and and antibottom quark in $e^+e^-$ annihilation
where the condition is fairly well satisfied. Perturbative $Q\bar Q$
systems are the ones that are best understood. They allow for a rather
precise quantitative description from first principles QCD,
i.e. without  a sizeable model-dependence. This makes perturbative
$Q\bar Q$ systems in principle 
an ideal tool to determine QCD parameters such as the heavy quark mass
$m_Q$ and the strong coupling $\alpha_s$. 

In this talk I review the recent achievements in the
description of perturbative non-relativistic $Q\bar Q$ systems using
the concept of effective field theories. These developments have
allowed a systematic determination of corrections of order $v^2$ with
respect to the non-relativistic limit for heavy quark production (or
annihilation) in $e^+e^-$ annihilation and opened up the access to 
the determination of even higher order corrections. In
Sec.~\ref{sectionmotivation} I will briefly motivate why the concept
of effective field theories is so helpful, and in fact necessary to
understand perturbative $Q\bar Q$ systems quantitatively, and in
Sec.~\ref{sectionEFT} 
I discuss basics of the construction of an effective theory for
non-relativistic perturbative $Q\bar Q$ systems ignoring
non-perturbative effects altogether. In
Sec.~\ref{sectioncrosssection} the calculation of the total
$Q\bar Q$ production cross section in $e^+e^-$ annihilation close to
threshold at NNLO is sketched, while in Sec.~\ref{sectionmass} the
need to introduce heavy quark mass definitions that are adapted to the 
non-relativistic framework is explained. Section~\ref{sectionttbar}
contains the application of the NNLO cross section to the top mass
determination at a future lepton pair collider. In
Sec.~\ref{sectionconclusions} I briefly mention further applications
and give the conclusions.

\section{Why do we need an Effective Theory?}
\label{sectionmotivation}
The lowest order description of the perturbative $Q\bar Q$ pair in the
non-relativistic expansion is provided by the well-known Schr\"odinger
equation 
\begin{eqnarray}
\bigg(\,
-\frac{\mbox{\boldmath $\nabla$}^2}{m_Q}  
-\frac{4}{3}\frac{\alpha_s}{|\mbox{\boldmath $r$}|}
- E
\,\bigg)\,G(\mbox{\boldmath $r$},\mbox{\boldmath $r$}^\prime,E) & = &
 \delta^{(3)}(\mbox{\boldmath $r$}-\mbox{\boldmath $r$}^\prime) 
\,.
\label{SchroedingerLO}
\end{eqnarray}
Because all relevant scales are larger than $\Lambda_{\rm QCD}$, the
coupling is weak and the potential can be calculated perturbatively
(similar to QED). Equation~(\ref{SchroedingerLO}) provides a
resummation of the most singular terms $\propto (\alpha_s/v)^n$ in the
elastic  
$Q\bar Q\to Q\bar Q$ scattering amplitude for $v\to 0$ in full
QCD. There is nothing more we have to do in the non-relativistic
limit. However, a problem arises when we try to calculate relativistic
corrections. As an example, let us consider the Darwin term 
$\delta
H_D=\frac{4}{3}\frac{\alpha_s\pi}{m_Q^2}
\delta^{(3})(\mbox{\boldmath $r$})$. 
The first order corrections to the energy levels can be determined
trivially as $\delta E_n=\langle n|H_D|n\rangle\sim|\psi_n(0)|^2$,
using Rayleigh--Schr\"odinger
time-independent perturbation theory,
where $n$ stands for the appropriate set of quantum numbers. For the
description of production and decay rates, however, we also need
the corrections to the wave function at the origin. Formally the   
answer reads
\begin{equation}
\delta \psi_n(0) \,=\,\langle\,\mbox{\boldmath $0$}\,|
\sum\hspace{-5.5mm}\int\limits_{i\ne n}
 \frac{|\,i\,\rangle\,\langle \,i\,|}{E_n-E_i}\,H_D\,|\,n\,\rangle
\sim \lim_{E\to E_n}\bigg[\,
G(0,0,E)-\frac{|\psi_n(0)|^2}{E_n-E}
\,\bigg]\psi_n(0)
\,.
\end{equation}
Interestingly the expression is UV-divergent because $G(0,r,E)$ 
contains terms proportional to $1/r$ and $\ln r$ for $r\to 0$. (The
power divergence can already be found in the free Green 
function $G_{\rm free}(0,r,E)=\frac{m_Q}{4\pi r}e^{i p r}$, 
$p=\sqrt{E m_Q}$.) The same divergence also exists in the
energy levels, but in the next order of perturbation theory.

Obviously, the UV-divergence reflects the fact that the
non-relativistic Schr\"odinger equation is only an approximation,
which is valid
for small momenta of order $m_Q v$ and smaller. This means that the
summation over the intermediate states with very large energies of
order $m_Q$ and larger is incorrect. The traditional approach to avoid
this problem has been promoted by Bethe and Salpeter~\cite{Bethe1} and
simply avoids the non-relativistic approximation completely. This is
achieved by formulating the problem in terms of fully relativistic
functional equations of Green functions. The Bethe--Salpeter approach,
however, has two disadvantages: first, the corresponding equations are
quite difficult to solve (because it is up to the person who solves
them to find out what exactly has to be calculated at a specific order
in the non-relativistic expansion) and, second, there is no systematic
way to separate perturbative and non-perturbative effects.

\section{The Effective Field Theory Approach}
\label{sectionEFT}
The effective field theory (EFT) approach avoids this dilemma. The EFT
approach for non-relativistic perturbative $Q\bar 
Q$ pairs had already been initiated some time ago by Caswell and
Lepage~\cite{Caswell1}, but only recently has it been fully
elaborated. To construct the EFT one has to separate those physical
degrees of freedom (d.o.f.'s) that can become on-shell for the
non-relativistic $Q\bar Q$ pair from those that 
can only be off-shell. This separation is, of course, not unambiguous.
So, the separation between what we count as on-shell and off-shell
is dependent on our choice of the
regularization scheme. The idea is to integrate out the
off-shell d.o.f.'s, in much the same way as particles with masses that
are much larger then the available energies can be integrated out in a
given process.
In order to describe production and annihilation processes, it
is also necessary to carry out the same program for external
currents. Once the EFT Lagrangian is obtained one can derive 
equations of motion for Green functions involving the $Q\bar Q$
pair. Here, the UV-divergence 
mentioned before does not arise from the fact that the EFT 
is defined within a regularization 
scheme. The resulting dependences on the regularization parameter
are cancelled once the Wilson coefficients of interactions and the
external currents in the EFT (which contain the contributions from the
off-shell d.o.f.'s) are taken into account.

For the case of the non-relativistic $Q\bar Q$ pair this program is 
non-trivial because we have to distinguish between on- and
off-shell components of the single heavy quark d.o.f. in full QCD. The
same it true for the gluon\footnote{
We assume that all light quarks are massless, and we count them as
gluons.  
}
d.o.f. in full QCD. 
A very useful way to identify the  relevant d.o.f.'s for the
non-relativistic $Q\bar Q$ system is to identify the momentum regimes 
needed to set up an asymptotic expansion of Feynman diagrams (in full
QCD) that describe elastic $Q\bar Q\to Q\bar Q$ scattering or $Q\bar
Q$ production and annihilation for centre-of-mass energies very close
to $2 m_Q$. It is obvious that, in general, one cannot naively carry
out the expansion {\it before} integration, because the resulting terms
might not lead to finite results. As a simple mathematical example
we might consider that we want to
determine the expansion of the parameter integral 
$f(a)=\int_0^1 \arctan(1-x)/(x^2+a^2) dx$ for $a\to 0$. Naive
expansion before integration will lead to divergent expressions. Here
we have to distinguish between the regions where $x\sim a$ and 
$x\sim 1$. We might introduce a cutoff $\Lambda$ that separates the
regions such that $a\ll \Lambda\ll 1$, but we could also use an analytic 
regularization scheme. In the region $x\sim a$ we can now
conveniently expand the 
$\arctan$ for small $x$, and for $x\sim 1$ we can expand the whole
integral in $a$. The integrations
can now be done rather easily. Adding both contributions and expanding
properly in $\Lambda$ and $a$, the cutoff dependence cancels and leaves
us with the correct expansion in $a$.

The corresponding momentum regions relevant to a non-relativistic
$Q\bar Q$ pair are more involved, because there are three widely
separated scales one has to take into account: $m_Q$, $m_Q v$ and $m_Q
v^2$. The regions were first fully identified by Beneke and
Smirnov~\cite{Beneke1}: 
$(k^0,{\mbox{\boldmath $k$}})\sim(m_Q,m_Q)$ (``hard''),
$(k^0,{\mbox{\boldmath $k$}})\sim(m_Q v,m_Q v)$ (``soft''),
$(k^0,{\mbox{\boldmath $k$}})\sim(m_Q v^2,m_Q v)$ (``potential'') and
$(k^0,{\mbox{\boldmath $k$}})\sim(m_Q v^2,m_Q v^2)$ (``ultrasoft'').
The energy and momentum components can scale differently with $v$
because the choice of a c.m.~system for the $Q\bar Q$ pair
breaks Lorentz invariance. All these regions can be populated by quark
and gluonic d.o.f.'s. For the $Q\bar Q$ system, where the available
energy is of order $m_Q v^2$, only quarks in the potential regime have
a chance to be on-shell. For gluons this can only happen in the
ultrasoft regime. In order to get to the EFT we have to integrate out
all the other d.o.f.'s. There have been different attempts and
philosophies to achieve this. Pineda and Soto~\cite{Pineda1} have proposed a
two-step procedure, where first the d.o.f.'s in the hard regime are
integrated out at the scale $m_Q$, leading to ``non-relativistic QCD''
(NRQCD), the theory formulated in~\cite{Caswell1}. In a second step,
the other off-shell d.o.f.'s are integrated out at a scale of order $m_Q
v$, leading to ``potential NRQCD'' (pNRQCD). 
This two-step procedure has the feature that it contains two separate
matching scales. Manohar and collaborators~\cite{Manohar1,Manohar2}
propose a single-step procedure where the  
off-shell d.o.f.'s are integrated out at the scale $m_Q$, except for
soft gluons. The resulting EFT is called ``velocity NRQCD''
(vNRQCD). The crucial issue of vNRQCD is that it is regulated by a
regularization velocity $\nu$ in dimensional regularization. This
regularization allows the scaling of the 
boundaries of the different momentum regions with $v$ at the same
time. A common feature of both EFTs is
that their Lagrangians contain four-quark $(Q\bar Q)(Q\bar Q)$
interactions that depend non-locally on the quark three momenta, but
locally on the quark energies.
These interactions represent a generalization of what is called
``potential'' in the Schr\"odinger equation~(\ref{SchroedingerLO}).
We note that, in the EFT for perturbative $Q\bar Q$ pairs, these
potentials are purely short-distance objects that can be calculated
perturbatively and do, in contrast to potential models, not contain
any confining contributions. 

Setting aside the problem of properly resumming logarithms of 
the velocity~\cite{resummationlogs}
(which is acceptable because the corresponding logarithms are not
dominant numerically for the practical applications involving $t\bar
t$ and $b\bar b$ pairs) both formulations of the EFT agree on how to
calculate the relativistic corrections to $Q\bar Q$ production. 
From now on, I will therefore only talk about ``the EFT for
perturbative non-relativistic $Q\bar Q$ pairs''. 
Based on the velocity scaling of the different momentum regimes, it is 
possible to unambiguously determine to which power in $v$ 
an individual interaction in the Lagrangian of the EFT scales. 
This ``power counting'' allows an unambiguous identification of the 
interactions (and the contributions in their Wilson coefficients)
that have to be taken into account to describe a process involving the
non-relativistic $Q\bar Q$ pair at a certain order in $v$ (i.e. in the 
non-relativistic expansion). One well-known fact that can be 
derived from formal power counting is that the Coulomb interaction
scales with the same power in $v$ as the kinetic energy, i.e. the
Coulomb interaction cannot be treated as a perturbation.
As a particularly important application of the power counting, one can
also show that the interaction of 
the quark with ultrasoft gluons is suppressed by $v^3$ with respect to
the non-relativistic limit described in the Schr\"odinger
equation~(\ref{SchroedingerLO}). This means that up to NNLO\footnote{
The non-relativistic limit described in the Schr\"odinger
equation~(\ref{SchroedingerLO}) is usually called ``leading-order'' (LO);
corrections of order $v^n$ with respect to the non-relativistic limit
are called N$^n$LO.
} 
we can
simply neglect the effects of ultrasoft gluons---a great
simplification because ultrasoft gluons cause
Lamb-shift-type retardation corrections that are quite difficult to
calculate. For $Q\bar Q$ production they have never been fully
calculated to date; only partial results exist so
far~\cite{ultrasoft}. 
Another important issue is that the EFT gives an exact definition for
the energy $E$ in Eq.~(\ref{SchroedingerLO}), which is particularly
important for the determination of heavy quark masses from
experimental data.

\section{The Cross Section $\sigma(e^+e^-\to Q\bar Q)$ Close to 
Threshold at NNLO}
\label{sectioncrosssection}
To determine the total $Q\bar Q$ production cross section in $e^+e^-$
annihilation in the non-relativistic region at NNLO
in the EFT, we start from the corresponding expressions in full QCD
($R\equiv\sigma(e^+e^-\to Q\bar Q)/\sigma(e^+e^-\to\mu^+\mu^-)$):\footnote{
For simplicity we consider only the photon-mediated cross
section.
}
\begin{eqnarray}
R(q^2) & = &
\frac{4\,\pi\,e_Q^2}{q^2}\,\mbox{Im}\,\bigg[\,
-i\,\int\!d^4x\,\,e^{i\,q.x}\,
  \langle\, 0\,| T\,j_\mu(x) \, j^{\mu}(0)\, |\, 0\,\rangle\,\bigg]
\nonumber\\[2mm] & \equiv &
\frac{4\,\pi\,e_Q^2}{q^2}\,\mbox{Im}\,\Big[\,-i\,
\langle\, 0\,| T\, \tilde j_\mu(q) \,
 \tilde j^{\mu}(-q)\, |\, 0\,\rangle\,\Big]
\,,
\label{crosssectioncovariant}
\end{eqnarray}
where $e_Q$ is the heavy-quark electric charge and $\sqrt{q^2}$
the c.m. energy; $\tilde j_\mu(\pm q)$ are the
electromagnetic currents that produce and annihilate the $Q\bar Q$
pair with c.m. energy $\sqrt{q^2}$. In the EFT these (external)
currents are replaced by a sum of 
${}^3\!S_1$ EFT currents with increasing dimension. For NNLO we need
currents up to dimension $8$ ($i=1,2,3$):
\begin{eqnarray}
\tilde j_i(q) & = & c_1\,\Big({\tilde \psi}^\dagger \sigma_i 
\tilde \chi\Big)(q) -
\frac{c_2}{6\,m_Q^2}\,\Big({\tilde \psi}^\dagger \sigma_i
(\mbox{$-\frac{i}{2}$} 
\stackrel{\leftrightarrow}{\mbox{\boldmath $D$}})^2
 \tilde \chi\Big)(q) + \ldots
\,,
\nonumber\\[2mm]
\tilde j_i(-q) & = & c_1\,\Big({\tilde \chi}^\dagger \sigma_i 
\tilde \psi\Big)(-q) -
\frac{c_2}{6\,m_Q^2}\,\Big({\tilde \chi}^\dagger \sigma_i
(\mbox{$-\frac{i}{2}$} 
\stackrel{\leftrightarrow}{\mbox{\boldmath $D$}})^2
 \tilde \psi\Big)(-q) + \ldots 
\,,
\label{currentexpansion}
\end{eqnarray}
where the Pauli spinors $\psi$ and $\chi$ describe the heavy quark and
antiquark fields in the EFT, respectively.
The constants $c_1$ and $c_2$ are Wilson coefficients that contain a
perturbative series in $\alpha_s$ originating from integrating out the
off-shell d.o.f.'s. At NNLO $c_1$ has to be calculated at order
$\alpha_s^2$ (Refs.~\cite{Hoang1,BenekeMelnikov1}) and contains
only contributions from hard d.o.f.'s. For $c_2$ 
the Born expression is sufficient. The dependence of $c_1$ on the
regularization parameter is not written out. We note that only the spatial
components of the currents contribute at the NNLO level.
Using the expansions~(\ref{currentexpansion}) we arrive at the 
following expression for the cross section at NNLO
\begin{eqnarray}
R_{\mbox{\tiny NNLO}}^{\mbox{\tiny thr}}(q^2) & = &
\frac{\pi\,e_Q^2}{m_Q^2}\,C_1\,
\mbox{Im}\Big[\,{\cal{A}}_1(q^2)\,\Big] 
- \frac{4 \, \pi\,e_Q^2}{3 \,m_Q^4}\,
C_2\,\mbox{Im}\Big[\,{\cal{A}}_2(q^2)\,\Big]
+ \ldots 
\,,
\label{crosssectionexpansion}
\end{eqnarray}
where
\begin{eqnarray}
{\cal{A}}_1(q^2) & = & i\,\langle \, 0 \, | 
\, ({\tilde\psi}^\dagger \vec\sigma \, \tilde \chi)(q)\,
\, ({\tilde\chi}^\dagger \vec\sigma \, \tilde \psi)(-q)\,
| \, 0 \, \rangle
\,,
\label{A1def}
\\[2mm]
{\cal{A}}_2(q^2) & = & \mbox{$\frac{1}{2}$}\,i\,\langle \, 0 \, | 
\, ({\tilde\psi}^\dagger \vec\sigma \, \tilde \chi)(q)\,
\, ({\tilde\chi}^\dagger \vec\sigma \, (\mbox{$-\frac{i}{2}$} 
\stackrel{\leftrightarrow}{\mbox{\boldmath $D$}})^2 \tilde \psi)(-q)\,
+ \mbox{h.c.}\,
| \, 0 \, \rangle
\,,
\label{A2def}
\end{eqnarray}
are EFT current--current correlators and $m_Q$ is the heavy quark pole
mass. Using the EFT equations of motion for the heavy quark fields 
${\cal{A}}_2$ can be related directly to ${\cal{A}}_1$:
\begin{equation}
{\cal{A}}_2 = m_Q\,E\,{\cal{A}}_1
\,.
\end{equation}
The current--current correlators are also 
regularization-parameter-dependent and, in fact cancel the
regularization-parameter dependence 
of $C_{1/2}$. ${\cal{A}}_1$ could be calculated from Feynman diagrams in
the EFT. However, even in LO we would have to resum an infinite number
of them (and we would have to carry out this resummation at NNLO). It is
more convenient to consider the four-point function 
($p\equiv(\sqrt{q^2},{\mbox{\boldmath $0$}}$))
\begin{eqnarray}
\tilde G(k,k^\prime,q^2)
\, = \,
\Big\langle \, 0 \, \Big| 
\,\Big({\tilde\psi}^\dagger(\frac{p}{2}+k)
       \tilde\chi(\frac{p}{2}-k)\Big)\,
\,\Big({\tilde\chi}^\dagger(-\frac{p}{2}+k^\prime)
       \tilde \psi(-\frac{p}{2}-k^\prime)\Big)\,
\Big| \, 0 \, \Big\rangle
\,,
\nonumber
\\
\label{A1Greenfunctionrelation}
\end{eqnarray}
that describes off-shell elastic scattering of a $Q\bar Q$ pair in the
EFT with c.m. energy $\sqrt{q^2}$. Integrating $\tilde G$
over $k$ and $k^\prime$ we arrive at ${\cal{A}}_1(q^2)$ (up to
a trivial factor). Upon integration over $k_0$ and $k^\prime_0$ the
equation of motion of $\tilde G$ at NNLO is just the Schr\"odinger
equation~(\ref{SchroedingerLO}) modified by NNLO corrections.:
\begin{eqnarray}
\lefteqn{
\bigg[\,
 \frac{\mbox{\boldmath $k$}^2}{M_t} - 
\frac{\mbox{\boldmath $k$}^4}{4M_t^3}
-(\sqrt{q^2} - 2\,m_Q -2\,\delta m_Q)
\,\bigg)
\,\bigg]\,
\tilde G(\mbox{\boldmath $k$},\mbox{\boldmath $k$}^\prime;q^2)
}
\nonumber\\[2mm] & &
 \,+\, 
\int\frac{d^3 \mbox{\boldmath $p$}}{(2\,\pi)^3}\,
\tilde V(\mbox{\boldmath $k$},\mbox{\boldmath $p$})\,
\tilde G(\mbox{\boldmath $p$},\mbox{\boldmath $k$}^\prime;q^2)
\, = \, 
(2\,\pi)^3\,\delta^{(3)}(\mbox{\boldmath $k$}-\mbox{\boldmath $k$}^\prime) 
\,.
\label{NNLOSchroedinger}
\end{eqnarray}
All terms displayed in Eq.~(\ref{NNLOSchroedinger}) are of order 
$v^2\sim\alpha_s^2$ (LO) and higher.
The potential $\tilde V$ contains the Coulomb potential shown in
Eq.~(\ref{SchroedingerLO}), its one- and two-loop
corrections~\cite{Schroeder1}, the order $\alpha_s/m_Q^2$ Breit--Fermi
potential known from positronium, and an order $\alpha_s^2/m_Q$
non-Abelian potential. This equation can be solved either
perturbatively, starting from the known LO Coulomb solution, or
exactly, using numerical methods. 
It is important that this calculation is carried out in 
exactly the same regularization scheme as was used to determine
the constants $C_{1/2}$.
As mentioned before, we obtain  ${\cal{A}}_1(q^2)$
by integrating 
$\tilde G(\mbox{\boldmath $k$},\mbox{\boldmath $k$}^\prime;q^2)$ over 
$\mbox{\boldmath $k$}$ and $\mbox{\boldmath $k^\prime$}$. From the EFT
we can also uniquely determine the exact definition of what is $E$
in Eq.~(\ref{SchroedingerLO}): in the pole mass scheme we find $\delta 
m_Q=0$ in Eq.~(\ref{NNLOSchroedinger}).

\section{The Heavy Quark Mass}
\label{sectionmass}
The pole mass definition seems to be the natural choice to formulate
the non-relativistic effective theory that describes the $Q\bar Q$
dynamics close to threshold. After all, the heavy quark pole mass is
gauge-invariant and IR-finite~\cite{Kronfeld1}. 
Moreover, in the pole mass 
scheme the equation of motion for the non-relativistic $Q\bar Q$ pair
has the simple form of Eq.~(\ref{SchroedingerLO}) (and also of
(\ref{NNLOSchroedinger}) for $\delta m_Q=0$), which is well known
from the non-relativistic problems in QED. Intuition also seems to
favour the pole mass definition, because close to threshold the heavy
quarks only have a very small virtuality of order $m_Q^2 v^2$, i.e. they
are very close to the mass shell. However, it is known
that the use of the pole mass can lead to (artificially) large high 
order corrections, because of its strong sensitivity to small
momenta~\cite{renormalon}. Quantitatively, the pole mass is ambiguous
to an amount of order $\Lambda_{\rm QCD}$. Thus the use of the pole
mass is unfavourable if we intend to use the calculation for the
$Q\bar Q$ cross section at NNLO for a precise determination of the
heavy quark 
mass. [At this point I emphasize that the notion of the ``heavy quark 
mass'' is not a physical one, because we can only define quark masses
as parameters in the QCD Lagrangian, not as physical quantities.]
In fact, the first theoretical calculations of the NNLO cross section
for $t\bar t$ production close to 
threshold~\cite{Hoang2,Melnikov1,Yakovlev1}
were carried out 
in the pole mass scheme; they seemed to indicate a rather large
instability of the location where the $t\bar t$ cross section at
threshold rises, because this rise did not seem to converge when
comparing LO, NLO and NNLO calculations. (See Ref.~\cite{Hoang3} for a  
synopsis of all calculations that have been done
for the total $t\bar t$ cross section at NNLO.)
The situation was similar for the first determinations of the
bottom quark mass from moments of the total $b\bar b$ cross section at
NNLO, where the uncertainties were quite 
large~\cite{Hoang4,Melnikov2} if the pole mass was employed and the
latter were estimated conservatively.  

The bottom line is that we should use a heavy quark mass definition
that has an infrared sensitivity much smaller than the pole
definition. In principle, the $\overline{\mbox{MS}}$ mass definition
might be an ideal candidate because its definition does not depend on
infrared physics. However, using the $\overline{\mbox{MS}}$ mass would
mean that $\delta m_Q=m_Q^{\rm pole}-\overline{m}_Q$ is of
order $\alpha_s$ and parametrically larger than any other term in the
Schr\"odinger equation~(\ref{NNLOSchroedinger}). This means that the 
$\overline{\mbox{MS}}$ mass formally breaks the non-relativistic power
counting. In order to find an adequate ``threshold mass'' definition,
it helps to have a closer look at Eq.~(\ref{NNLOSchroedinger}).
Technically, the instability observed in the pole mass scheme
originates from the potential $\tilde V$. It can be shown that the 
potential causes large corrections from momenta smaller than $m_Q
\alpha_s$ at large orders of perturbation theory.
This can be visualized by considering the small momentum contribution
to the heavy quark potential in configuration space representation for 
distances of the order of the inverse Bohr radius $1/m_Q\alpha_s$:
\begin{eqnarray}
\bigg[\,V(r\approx 1/M_t\alpha_s)\,\bigg]^{\rm IR}
& \sim & 
\int\limits^{|\mbox{\scriptsize\boldmath $q$}|<\mu\ll M_t\alpha_s}
\frac{d^3\mbox{\boldmath $q$}}{(2\pi)^3}\,
\tilde V_c(\mbox{\boldmath $q$})\,
\exp(i\,\mbox{\boldmath $q$}\mbox{\boldmath $r$})
\\[2mm] & = &
\int\limits^{|\mbox{\scriptsize\boldmath $q$}|<\mu\ll M_t\alpha_s}
\frac{d^3\mbox{\boldmath $q$}}{(2\pi)^3}\,
\tilde V_c(\mbox{\boldmath $q$}) \, + \, \ldots
\,,
\label{lowmomV}
\end{eqnarray}
where $\tilde V_c$ is the Coulomb potential in momentum space
representation. At large orders of perturbation theory the RHS of
Eq.~(\ref{lowmomV}) grows asymptotically like $-\mu\,\alpha_s^n
n!$ (Ref.~\cite{Aglietti1}), where $\mu$ is the scale of $\alpha_s$. On
the other hand, it can 
be shown that the total static energy $2 m_Q^{\rm pole} + V_c(r)$ does
not contain these large corrections~\cite{Beneke2,Hoang5,Uraltsev1}. 
Therefore, an adequate mass definition can be found by reabsorbing the
$r$-independent small-momentum contribution of the Coulomb potential 
into the heavy quark mass. This procedure is not unique. Several
different mass definitions have been suggested in the
literature~\cite{Beneke2,Czarnecki1,Hoang6} that fulfil this task.  
Let me mention only the 
1S mass~\cite{Hoang6} in some detail. The 1S mass is
defined as one half of the mass of the perturbative contribution of a
fictitious $n=1$, ${}^3S_1$ $Q\bar Q$  bound state, assuming that the
quarks are stable:
\begin{eqnarray}
m_Q^{\rm 1S}-m_Q^{\rm pole} & = & \frac{1}{2}\,
\langle 1\,{}^3S_1 | {\cal{H}} |  1\,{}^3S_1 \rangle
 \, = \,
- \frac{2}{9}\,\alpha_s^2\,m_Q 
\, + \, \ldots
\,,
\label{1Sdef}
\end{eqnarray}
where ${\cal{H}}$ is the non-relativistic perturbative Hamiltonian.
The NNLO expression for the RHS of Eq.~(\ref{1Sdef}), which is not
displayed here, was first calculated in Ref.~\cite{Pineda2}. 
The 1S mass is less sensitive to infrared momenta than the pole mass
(its ambiguity is only of order $\Lambda_{\rm QCD}^2/m_Q$) and, at the
same time, does not break the non-relativistic power counting.
 
In order to relate the 1S mass to the $\overline{\mbox{MS}}$ mass,
some care has to be taken, because it is not a priori clear how to
combine the corresponding perturbative series. The guiding principle
is the proper cancellation of the asymptotically large perturbative
corrections contained in the perturbative relations of the 1S and the 
$\overline{\mbox{MS}}$ masses to the pole mass in each order of
perturbation theory. It can be shown that the
cancellation happens properly if a modified perturbative expansion,
called the ``upsilon expansion''~\cite{upsilon} is used. In the upsilon
expansion one has to consider terms of order $\alpha_s^{n}$ in
Eq.~(\ref{1Sdef}) as of order $\alpha_s^{n-1}$. Technically this
unusual prescription comes from the fact that the coefficient of the
order $\alpha_s^{n}$ term in Eq.~(\ref{1Sdef}) contains the contribution
$\sum_{i=0}^{n-2}\ln(1/\alpha_s)/i!$, which leads asymptotically to an
additional factor $1/\alpha_s$. The upsilon expansion
should be applied whenever the 1S mass is used outside the framework
of the non-relativistic problems. As an example, it has been
applied for the description of inclusive semileptonic B
decays~\cite{upsilon}, where it leads to a very well behaved
perturbative expansion.  

\section{Top-Antitop Production Close to Threshold}
\label{sectionttbar}
Top--antitop pair production at future lepton pair colliders is one
of the most important applications of the EFT for non-relativistic
$Q\bar Q$ pairs discussed in this talk. The $t\bar t$ production close to
threshold will provide a rich phenomenological environment where the
top quark can be studied in quite a unique manner.
Assuming the upcoming Linear Collider or a muon collider
will spend around a $100\,{\rm fb}^{-1}$ at the top--antitop threshold,
the top mass can be determined from the $t\bar t$ line shape with
experimental uncertainties below 50~MeV. The fact that the $t\bar t$
pair is produced in a colour singlet state is crucial to make this
possible. Even the top quark width might 
be extracted from the line shape with an uncertainty of around 20\%,
assuming that the theoretical uncertainties are small. Assuming that
the Higgs mass is around 100~GeV also a fairly good determination of
the top Yukawa coupling could be achieved. In addition,
from studies of distributions one might get  measurements of
anomalous top couplings and further measurements of the top width.
What makes the $t\bar t$ system so interesting theoretically is the
fact that its total width ($\Gamma_t\approx 1.5$~GeV) is much larger
than $\Lambda_{\rm QCD}$. Thus, the top quark decays before it
hadronizes, and the effects of non-perturbative effects should be
small. One can show that the effective velocity of the top quarks
close to 
threshold reads $v_{\rm eff}=|\sqrt{(E+i\Gamma_t)/m_Q}|$ rather than
$\sqrt{E/m_Q}$. This means that the hierarchy
$m_t\gg m_t v_{\rm eff}\gg m_t v_{\rm eff}^2>\Lambda_{\rm QCD}$ holds
even for c.m. energies that are extremely close to the threshold and
that non-perturbative effects should only represent relatively small
corrections.  

In view of this prospect, it is obvious that a careful analysis and
assessment of theoretical uncertainties in the
prediction of the total cross section is mandatory, in order to
determine whether the theoretical precision can meet the experimental
one. Within the last two and a half years, considerable progress has
been achieved in the calculation of NNLO corrections to the $t\bar t$
production process within the framework of the EFT for
non-relativistic $Q\bar Q$ pairs. For the total cross section the
complete set of NNLO QCD corrections have been 
calculated~\cite{Hoang2,Melnikov1,Yakovlev1,Beneke3,Nagano1,Penin1}. So
far, for the total cross section the complete electroweak effects have
only been implemented at NLO. The implementation up to NLO is quite easy:
the only thing to do is to replace 
$(\sqrt{q^2} - 2\,m_Q -2\,\delta m_Q)$ by 
$(\sqrt{q^2} - 2\,m_Q -2\,\delta m_Q)+i\Gamma_t$ in
Eq.~(\ref{NNLOSchroedinger}). The width has to be counted as of order 
$\alpha_s^2\sim v_{\rm eff}^2$ and cannot be treated as a
perturbation.
\begin{figure}[t!] 
\begin{center}
\leavevmode
\epsfxsize=2.8cm
\leavevmode
\epsffile[220 580 420 710]{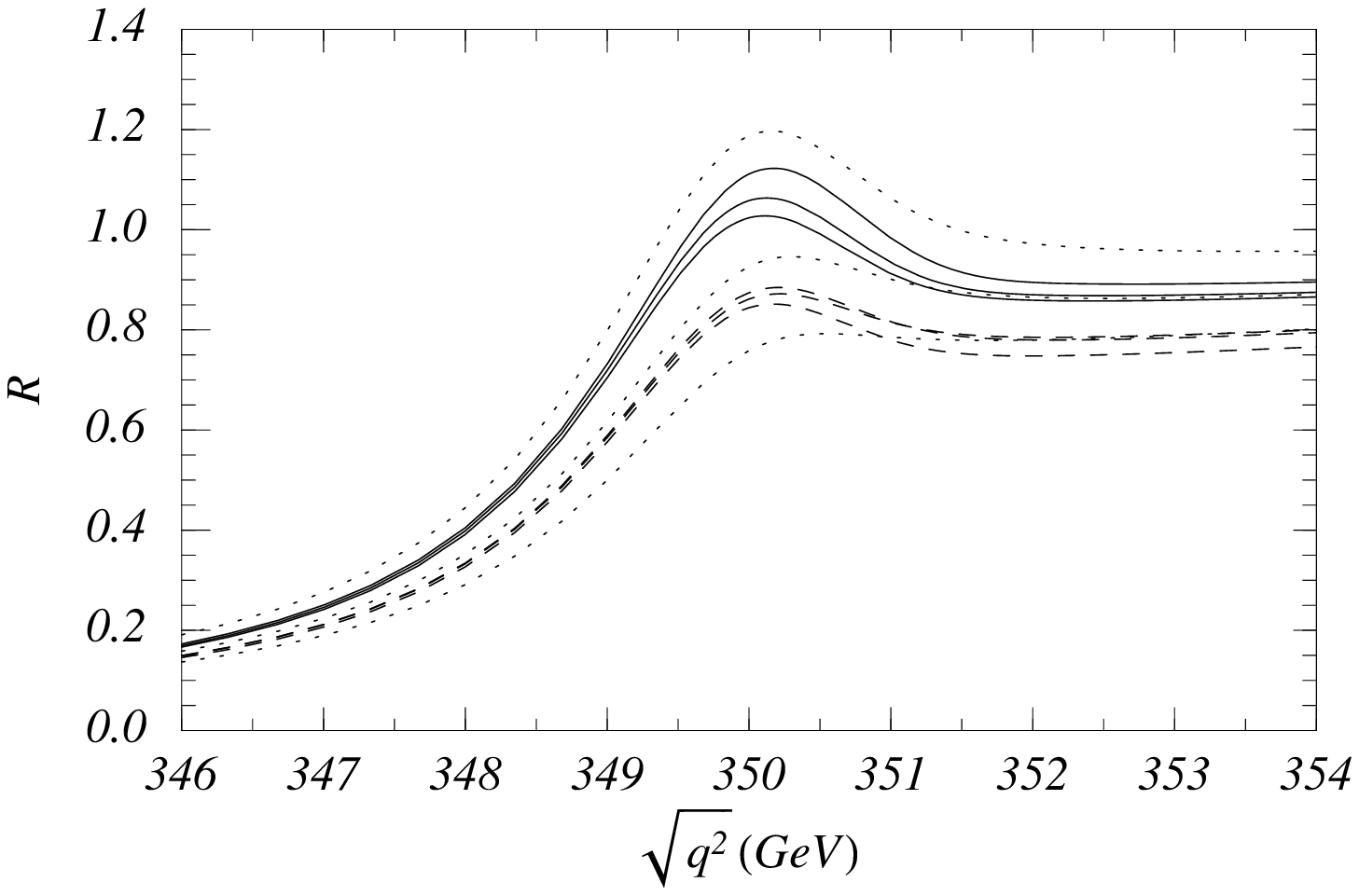}
\hspace{3cm}
\epsfxsize=2.8cm
\leavevmode
\epsffile[220 580 420 710]{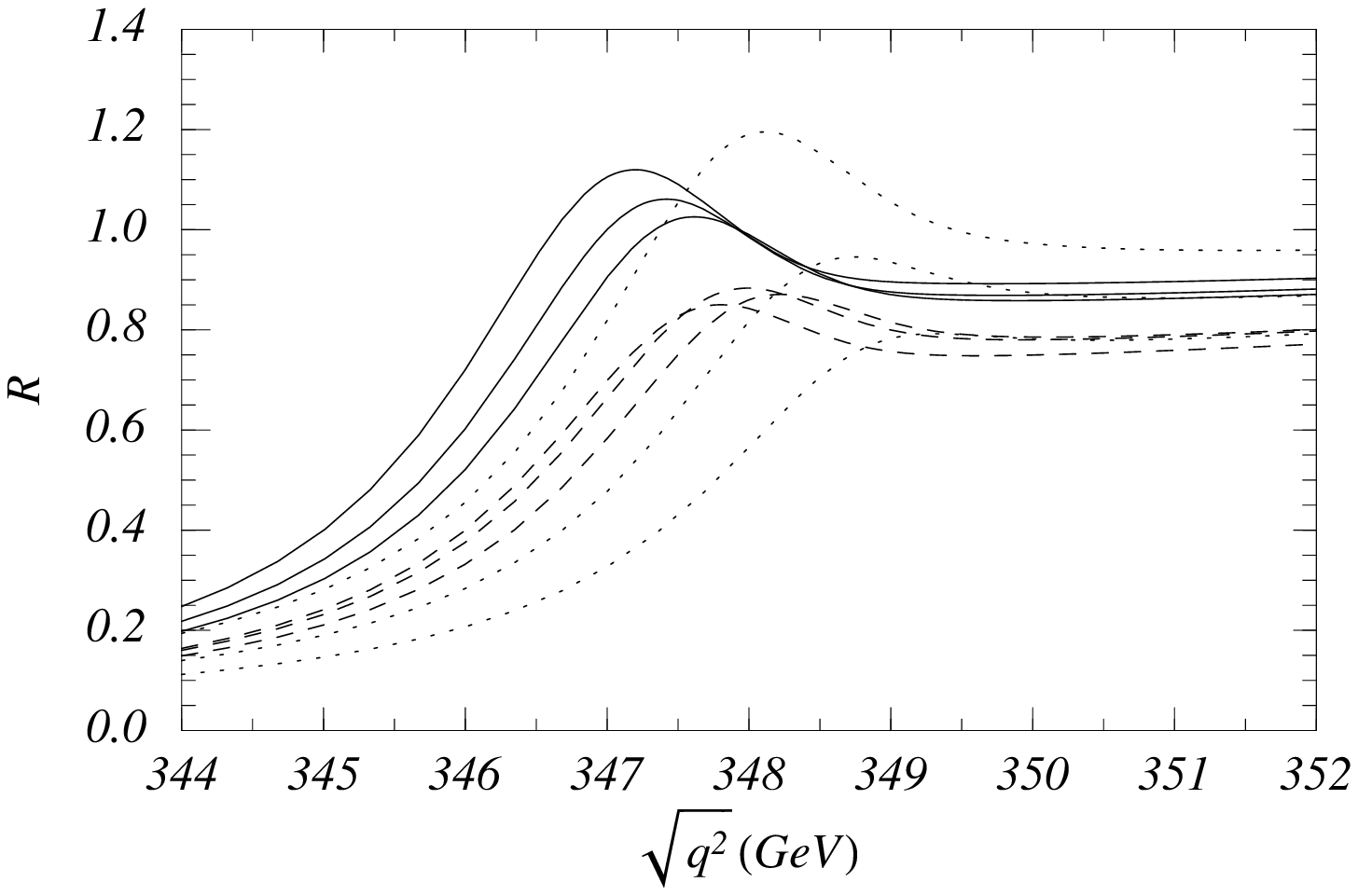}
 \end{center}
%
%
\vskip  1.7cm
 \caption{\label{figttbar}
The total photon-induced $t\bar t$ cross section divided by the point
cross section at the LC versus the c.m. energy in the threshold regime
at LO (dotted curves), NLO (dashed) and NNLO (solid) in the 1S
(left figure) and the pole (right figure) mass schemes for
$\alpha_s(M_Z)=0.118$ and $\mu=15$, $30$, $60$~GeV. The effects of
beam-strahlung and initial-state radiation have not been included. The
plots are taken from Ref.${}^{27}$.\mbox{\hspace{6cm}} 
}
\end{figure}
In the left picture of Fig.~\ref{figttbar} the result for the total
$t\bar t$ production cross section is shown for $m_t^{1S}=175$~GeV,
where dotted, dashed and solid curves represent LO, NLO and NNLO
results, respectively. As a comparison, the corresponding results are
shown in the pole mass scheme in the right picture of
Fig.~\ref{figttbar}. It is conspicuous that the normalization
uncertainties are rather large, at the level of 20\%. However,
realistic simulation studies~\cite{Peralta1} have shown that these
uncertainties do not severely affect the determination of a threshold
mass such as the 1S mass. A top mass determination with combined
theoretical and experimental uncertainties of order or smaller than
200~MeV seems realistic. 

\section{Other Applications and Conclusions}
\label{sectionconclusions}
There are a number of other applications of the EFT, which I will
only discuss briefly. As mentioned in the introduction, moments of the
total $b\bar b$ cross section in $e^+e^-$ annihilation can be used to
determine the bottom quark mass. These moments are defined as
\begin{equation}
P_n\equiv\int\frac{d q^2}{(q)^{2(n+1)}}
\frac{\sigma_{e^+e^-\to b\bar b+X}(q^2)}
{\sigma_{e^+e^-\to \mu^+\mu^-}(q^2)}
\,.
\end{equation} 
It can be shown that the effective velocity of the bottom quarks in
the $n$-th moments is $v_{\rm eff}=1/\sqrt{n}$, which means that 
$m_b> m_b v_{\rm eff}> m_b v_{\rm eff}^2>\Lambda_{\rm QCD}$ holds
as long as $n$ is not chosen larger than about 10. Determinations of the
bottom quark mass based on threshold mass definitions have been
carried out in Refs.~\cite{Melnikov2,Hoang7,Beneke4}. The results are
perfectly consistent and 
yield a value for the $\overline{\mbox{MS}}$ bottom mass $\overline
m_b(\overline m_b)$ of $4.2$~GeV, with an uncertainty of $60$--$80$~MeV.

It is clear that the non-relativistic EFT for heavy $Q\bar Q$ pairs
can also be applied to two-body systems where the binding is caused by
the electromagnetic forces. Apart from classic systems such as
hydrogen, positronium or muonium, where the use of effective field
theoretical methods has become increasingly popular within the last
two years, hadronic atoms are a quite interesting field of
application. Among the most interesting of these systems are the bound
state of two charged pions ($\pi^+\pi^-$), called pionium, or muonic
or kaonic hydrogen. A number of experiments such as DIRAC at CERN or
DEAR at DA{$\Phi$}NE are already running or are under way, with the aim
of measuring parameters of the chiral Lagrangian with unprecedented
precision in order to provide tests of our understanding of chiral
symmetry breaking. 

To conclude, one can say that the advent of EFT methods has
revived the interest of many theorists in non-relativistic two-body
systems. The strength of the EFT approach is that it represents a
``theory'' and not just a ``method''. It explicitly uses the concepts of
separation of scales, factorization and renormalization, and it
simplifies systematic calculations through a set of power counting
rules that can 
be applied before any calculation is actually started. This has made
non-relativistic heavy $Q\bar Q$ systems accessible to a wider public.
Today it is clear how to calculate NNNLO corrections to the
total $Q\bar Q$ cross section. This is a tremendous technical project,
but it is clear conceptually. The calculation of these corrections
might be important in obtaining a more realistic estimate of
theoretical uncertainties in the non-relativistic expansion for $t\bar
t$ and $b\bar b$ systems. 
An important conceptual development, which still has to be achieved, is
the proper implementation of the instability of a heavy quark such as
the top into the EFT framework. To study instability, the
non-relativistic framework seems to be a much better setting than the
full theory, because the non-relativistic EFT provides a natural 
separation of on- and off-shell heavy quark d.o.f.'s. 

\section*{Acknowledgements}
I thank Herbert Fried, Bernd M\"uller and Yves Gabellini for the
organization of this nice and informative conference.
This work is supported in part by the EU Fourth Framework Program
``Training and Mobility of Researchers'', Network ``Quantum
Chromodynamics and Deep Structure of Elementary Particles'', contract
FMRX-CT98-0194 (DG12-MIHT). 

\newpage

\sloppy
\raggedright
\def\app#1#2#3{{\em Act. Phys. Pol. }{\bf B #1} (#2) #3}
\def\apa#1#2#3{{\em Act. Phys. Austr.}{\bf #1} (#2) #3}
\def\lhc{Proc. LHC Workshop, CERN 90-10}
\def\npb#1#2#3{{\em Nucl. Phys. }{\bf B #1} (#2) #3}
\def\npps#1#2#3{{\em Nucl. Phys. Proc. Suppl.}{\bf #1} (#2) #3}
\def\nP#1#2#3{{\em Nucl. Phys. }{\bf #1} (#2) #3}
\def\plb#1#2#3{{\em Phys. Lett. }{\bf B #1} (#2) #3}
\def\prd#1#2#3{{\em Phys. Rev. }{\bf D #1} (#2) #3}
\def\pra#1#2#3{{\em Phys. Rev. }{\bf A #1} (#2) #3}
\def\pR#1#2#3{{\em Phys. Rev. }{\bf #1} (#2) #3}
\def\prl#1#2#3{{\em Phys. Rev. Lett. }{\bf #1} (#2) #3}
\def\prc#1#2#3{{\em Phys. Reports }{\bf #1} (#2) #3}
\def\cpc#1#2#3{{\em Comp. Phys. Commun. }{\bf #1} (#2) #3}
\def\nim#1#2#3{{\em Nucl. Inst. Meth. }{\bf #1} (#2) #3}
\def\pr#1#2#3{{\em Phys. Reports }{\bf #1} (#2) #3}
\def\sovnp#1#2#3{{\em Sov. J. Nucl. Phys. }{\bf #1} (#2) #3}
\def\sovpJ#1#2#3{{\em Sov. Phys. JETP Lett. }{\bf #1} (#2) #3}
\def\jl#1#2#3{{\em JETP Lett. }{\bf #1} (#2) #3}
\def\jet#1#2#3{{\em JETP }{\bf #1} (#2) #3}
\def\zpc#1#2#3{{\em Z. Phys. }{\bf C #1} (#2) #3}
\def\ptp#1#2#3{{\em Prog.~Theor.~Phys.~}{\bf #1} (#2) #3}
\def\nca#1#2#3{{\em Nuovo~Cim.~}{\bf #1A} (#2) #3}
\def\ap#1#2#3{{\em Ann. Phys. }{\bf #1} (#2) #3}
\def\hpa#1#2#3{{\em Helv. Phys. Acta }{\bf #1} (#2) #3}
\def\ijmpA#1#2#3{{\em Int. J. Mod. Phys. }{\bf A #1} (#2) #3}
\def\ZETF#1#2#3{{\em Zh. Eksp. Teor. Fiz. }{\bf #1} (#2) #3}
\def\jmp#1#2#3{{\em J. Math. Phys. }{\bf #1} (#2) #3}
\def\yf#1#2#3{{\em Yad. Fiz. }{\bf #1} (#2) #3}
\def\epjc#1#2#3{{\em Eur. Phys. J. C }{\bf #1} (#2) #3}
\def\epjdc#1#2#3{{\em Eur. Phys. J. direct C }{\bf #1} (#2) #3}


\begin{thebibliography}{99}
\bibitem{Brambilla1}
For a discussion on the situation where $\Lambda_{\rm QCD}$ is
parametrically larger than $m_Q v^2$, see the talk by N. Brambilla,
these proceedings.
\end{thebibliography}
\end{document}